\documentclass[a4paper,fleqn,usenatbib,useAMS]{mnras}

\usepackage[T1]{fontenc}
\usepackage{ae,aecompl}

\usepackage{graphicx}	
\usepackage{amsmath}	
\usepackage{amssymb}	

\title[Modeling transitional disks with Herschel]{Constraining the properties of transitional disks in Chamaeleon~I
  with Herschel\thanks{Herschel is an ESA space observatory with science instruments provided by European-led
    Principal Investigator consortia and with important participation from NASA.}}

\author[\'A. Ribas et al.]{
\'A. Ribas,$^{1}$\thanks{E-mail: \href{aribas@bu.edu}{aribas@bu.edu}}
H. Bouy,$^{2}$
B. Mer\'in,$^{3}$
G. Duch\^ene,$^{4,5}$
I. Rebollido,$^{6}$
C. Espaillat,$^{1}$
\newauthor{
and C.Pinte$^{5,7}$}
\\
$^{1}$Department of Astronomy, Boston University, 725 Commonwealth Avenue, Boston, MA 02215, USA\\
$^{2}$Centro de Astrobiolog\'{i}a, INTA-CSIC, P.O. Box - Apdo. de correos 78, Villanueva de la Ca\~nada Madrid 28691, Spain\\
$^{3}$European Space Astronomy Centre, ESA, P.O. Box, 78, 28691 Villanueva de la Ca\~{n}ada, Madrid, Spain\\
$^{4}$Astronomy Department, University of California, Berkeley, CA 94720-3411, USA\\
$^{5}$Univ. Grenoble Alpes, IPAG, F-38000 Grenoble, France\\
$^{ }$ CNRS, IPAG, F-38000 Grenoble, France\\
$^{6}$Universidad Aut\'onoma de Madrid, Departamento de F\'isica Te\'orica M\'odulo 15, 28049 Madrid, Spain\\
$^{7}$UMI-FCA, CNRS/INSU France (UMI 3386), and Departamento de Astronom{\'\i}a, Universidad de Chile, Casilla 36-D Santiago, Chile
}

\date{Accepted 2016 February 9. Received 2016 January 11; in original form 2015 November 11}

\pubyear{2015}

\begin{document}
\label{firstpage}
\pagerange{\pageref{firstpage}--\pageref{lastpage}}
\maketitle

\begin{abstract}

  Transitional disks are protoplanetary disks with opacity gaps/cavities in their dust distribution,
  a feature that may be linked to planet formation. We perform Bayesian modeling of the three
  transitional disks SZ~Cha, CS~Cha and T25 including photometry from the \emph{Herschel Space
    Observatory} to quantify the improvements added by these new data. We find disk dust masses
  between 2$\times$10$^{-5}$ and 4$\times$10$^{-4}$\,M$_\odot$ and gap radii in the range of
  7-18\,AU, with uncertainties of $\sim$\,one order of magnitude and $\sim$\,4\,AU,
  respectively. Our results show that adding \emph{Herschel} data can significantly improve these
  estimates with respect to mid-infrared data alone, which have roughly twice as large uncertainties
  on both disk mass and gap radius. We also find weak evidence for different density profiles with
  respect to full disks. These results open exciting new possibilities to study the distribution of
  disk masses for large samples of disks.

\end{abstract}



\section{Introduction} \label{sec:intro}

Transitional disks (TDs) are one of the main research topics within the current paradigm of planet formation: these
are protoplanetary disks with signatures of cavities and/or gaps in their dust distribution, which could be
directly linked to forming planets \citep[see][for an updated review on this field]{Espaillat2014}. For this
reason, a proper characterization of these systems could set strong constrains on the conditions under which
planets come to be. However, these holes could also be produced by other mechanisms such as photoevaporation, dust
growth and settling towards the disk mid-plane, dynamic clearing by (sub)stellar companions, or a combination of
several of these \citep[e.g.][]{Ireland2008, Birnstiel2012, Alexander2014, Espaillat2014}.  Despite the large
number of studies of TDs covering the whole wavelength domain with various observing techniques (e.g., photometry,
spectroscopy, polarimetry, or interferometry), several questions remain open, such as their real
connection with planet formation, whether every protoplanetary disk goes through a transitional phase, or the main
processes behind their evolution.

The Spectral Energy Distributions (SEDs) of full, optically thick circumstellar disks (Class-II)
have significant infrared (IR) excess with respect to the photospheric level from the near-infrared
($\sim$\,NIR, 1-5\,$\mu$m) to millimeter wavelengths \citep{Williams2011}. In contrast, SEDs of TDs
normally display a very distinctive shape, with small or no NIR excess, but with larger excess at
mid-infrared ($\sim$\,MIR, 5-50\,$\mu$m) and far-infrared (FIR, 50\,$\mu m$ and longer)
wavelengths. This lack of NIR emission is usually attributed to dust depleted inner regions in the
disk, and the location and shape of this change in the SED (around $\sim$10\,$\mu$m) can be used to
partially characterize the gap structure \citep[e.g.][]{Espaillat2010, Espaillat2011}. Objects with
small NIR excess and significant MIR and FIR excesses are known as pre-transitional disks
\citep[pre-TDs,][]{Espaillat2007b} and are thought to be an intermediate stage between full disks
and TDs: these have some optically thick dust in their inner region, separated from the outer disk
by an optically thin gap \citep[see e.g.][]{Espaillat2014}. Given their possible connection with
planet formation, (pre-)TDs have been extensively studied and modeled in the past with MIR spectra
from the the IRS instrument \citep{IRS} on board the \textit{Spitzer Space Telescope}, which yielded
thousands of IR spectra of circumstellar disks between 5-38\,$\mu$m and provided detailed
information about their dust composition and the morphology of the inner regions
\citep[e.g.][]{Kim2009, Merin2010, Espaillat2011, Furlan2011}. However, several parameters such as
the disk flaring or mass remain poorly or completely unconstrained with SED modeling and MIR data
alone. A more in-depth knowledge of TDs can be achieved with complementary (sub)mm data, where the
disk becomes mostly optically thin and the flux can be related to the disk mass
\citep[e.g.][]{Andrews2005, Najita2007, Andrews2013}. The advent of the \emph{Herschel Space
  Observatory}\citep{Herschel} produced a large number of FIR observations of protoplanetary and TDs
in the 70-500\,$\mu$m regime, providing us with information at larger wavelengths that can be used
to better constrain some parameters of TDs.

In this paper, we add \emph{Herschel} data to the modeling of three (pre-)TDs: SZ~Cha (a
  pre-TDs), CS~Cha (a binary system surrounded by a disk with a large cavity), and T25 (a TD with no
  known companion). These objects belong to the Chamaeleon~I region, located at 160-180\,pc from the
  Sun \citep[][and references therein]{Whittet1997}, and with an age estimate of $\sim$\,2\,Myr
  \citep{Luhman2008a}.  Given its proximity, Chamaeleon~I has been an usual target for
  star-formation and stellar population studies \citep[e.g.,][]{Luhman2007,Luhman2008a,
    Belloche2011}, which identified more than 200 YSOs in the region. It was also one of the clouds
  observed by \emph{Herschel}, and some studies have already provided a \emph{Herschel} view of its
  YSO population \citep{Winston2009}, disks around low-mass stars \citep{Olofsson2013}, and TDs
  \citep{Ribas2013, Rodgers-Lee2014}. Here, we focus on the impact of \emph{Herschel} data in
  different parameters of these disks obtained from SED modeling, by combining radiative transfer
  modeling with Markov Chain Monte Carlo (MCMC) methods to perform a Bayesian analysis of their
  properties. Sect.~\ref{sec:observations} describes the sample and data used. The modeling
procedure can be found in Sect.~\ref{sec:modeling}, and the results of the process are described in
Sect.~\ref{sec:results}. Finally, we discuss the implications of our analysis in
Sect.~\ref{sec:discussion}.

\section{Observations and Data Reduction} \label{sec:observations}

\subsection{The sample}

A total of 12 sources have been previously classified as TDs and pre-TDs in the Chamaeleon~I region.
Several of these targets have been modeled in detail, mostly based on their \emph{Spitzer} MIR spectra
\citep[e.g.][]{Kim2009}. Additionally, a number of studies have already explored \emph{Herschel} data of these
disks \citep[e.g.][]{Winston2012, Ribas2013,Rodgers-Lee2014}. To analyze the feasibility of estimating disk masses
using \emph{Herschel} data of (pre-)TDs, we selected three different objects in Chamaeleon~I:

\begin{itemize}

\item SZ~Cha, a pre-TD \citep{Kim2009}, 
\item CS~Cha, a disk with a gap surrounding a binary system \citep[binary separation of
  $\sim$\,3.5\,AU,][]{Nagel2012}, i.e., a circumbinary disk \citep[][]{Guenther2007, Nagel2012},
\item T25, a TD with no known companion \citep{Kim2009}.

\end{itemize}

These three objects were selected because they represent the main scenarios in
our current understanding of disk evolution: from a pre-TD (SZ~Cha), to objects
with clean opacity holes either caused by binaries (CS~Cha) or other mechanisms
(T25). For these targets, we used the stellar parameters in
\citet{Espaillat2011}, which are listed in
Table~\ref{tab:stellar_parameters}. In the case of CS~Cha, the orbital
  motion of the binary system may change the radiation received by different
  regions of the disk with time, and hence the emission from the inner disk
  varies with a similar period. \citet{Nagel2012} used two-star models to show
  that the variability produced by this effect in this case is only of
  $\sim$1\,\% at the 10\,$\mu$m peak, and we approximate the binary system by a
  single star (our photometric uncertainties and model noise are larger than
  this value, see Sect.s~\ref{sec:nearIR} and~\ref{sec:IRS}). We therefore adopt
  the spectral type provided in \citet{Luhman2004ChaI} for this target, which
  has also been used in previous modeling efforts \citep{Espaillat2007a,
    Kim2009, Manoj2011,Espaillat2011}, allowing for meaningful comparisons.

\begin{table*}
  \caption{Coordinates and stellar parameters used in this work for the considered sample of (pre-)transitional
    disks. Stellar parameters as in \citet{Espaillat2011}.}\label{tab:stellar_parameters}
  \begin{center}
    {\footnotesize
    \begin{tabular}{l c c c c c c c c}
      \hline
      \hline\rule{0mm}{3mm}Name & R.A.$_{J2000}$ & Dec.$_{J2000}$ & A$_V$ & SpT & T$_*$ & L$_*$ & M$_*$ & R$_*$ \\
                                & & & (mag) & & (K) & (L$_\odot$) & (M$_\odot$) & (R$_\odot$) \\
      \hline
      SZ Cha & 10:58:16.77 & -77:17:17.1 & 1.9 & K0 & 5250 & 1.9 & 1.4 & 1.7 \\
      CS Cha & 11:02:24.91 & -77:33:35.7 & 0.8 & K6 & 4205 & 1.5 & 0.9 & 2.3 \\
      T25 & 11:07:19.15 & -76:03:04.9 & 1.6 & M3 & 3470 & 0.3 & 0.3 & 1.5 \\
      \hline
    \end{tabular}
  }
  \end{center}
\end{table*}

\subsection{\emph{Herschel} data}

In \citet{Ribas2013} we presented \textit{Herschel} aperture photometry measurements of the TDs in
Chamaeleon~I. Given the inherent difficulties in determining whether a source is detected or
  not in the presence of conspicuous background, in that previous study we visually inspected the
  position of the known YSOs in the region \citep[see][for a complete description of the
  sample]{Ribas2013}. In this paper, we maintain this criterion, but also expand the analysis of T25
  which was previously undetected at 500\,$\mu$m (see below). Later, \citet{Rodgers-Lee2014}
identified a systematic discrepancy between the photometry in \citet{Ribas2013} and the one in
\citet{Winston2012}, which used the \emph{getsources} algorithm \citep{getsources}. We attribute
such discrepancies to the different map-making and source extraction algorithms used, but for the
sake of completeness we chose to re-process the corresponding data. Here, we describe the adopted
methods.

The Chamaelon~I region was observed by \textit{Herschel} as part of the Gould Belt Key Program
\citep{Andre2010}. Parallel mode observations from this program (OBSDIs: 1342213178, 1342213179)
provided PACS \citep{PACS} 70 and 160\,$\mu$m and SPIRE \citep{SPIRE} 250, 350 and 500\,$\mu$m maps
at a scan speed of 60\arcsec/s. Additional PACS 100 and 160 scan observations are also available in
the Gould Belt Key Program at a scan speed of 20\arcsec/s (OBSIDs: 1342224782, 1342224783). Although
with smaller coverage, this last set of observations is deeper and has a slower scan speed (hence a
smaller point spread function, PSF), so we chose to use them for the 160\,$\mu$m band instead of the
parallel mode data. We note that T25 is outside the smaller scan maps, and no 100\,$\mu$m
photometric measurement is available for this target. Its 160\,$\mu$m flux was therefore obtained
from the parallel mode data.

We used the Herschel Interactive Processing Environment \citep[HIPE,][]{Ott2010} version 12.1 to
process the maps of the region. We adopted the standard map-making algorithms used in the
\textit{Herschel} Science Archive, i.e., \textit{jscanam} for PACS maps \citep[a HIPE adaptation of
the \textit{Scanamorphos} software,][]{Scanamorphos}, and \textit{destriper} for SPIRE maps. In the
case of \textit{jscanam} we remove turnarounds with speeds 50\,\% lower or higher than the nominal
speed value, and we do not use the extended emission gain option for \textit{destriper}, as
recommended for point source photometry.

We estimated PACS fluxes with the \emph{AnnularSkyAperturePhotometry} task in HIPE. We adopted aperture radii of
15\,\arcsec, 15\,\arcsec, and 22\,\arcsec\, for PACS 70, 100 and 160 bands, respectively.  These apertures were
specifically selected after inspecting growth curves of each target in each band. The background was estimated
within annulus with inner and outer radii of 25\,\arcsec and 35\,\arcsec. We applied the corresponding aperture
corrections of 1.206, 1.222, and 1.372 \citep{Balog2014}.  Given the negligible color corrections for PACS for
temperatures above 20\,K (PACS Photometer - Color Corrections manual, version 1.0) and the uncertainty in
determining the slope of the SED close to the emission peak, we chose not to apply them for PACS bands. 
case they are significantly smaller than the adopted photometric uncertainties (see below).  For SPIRE, we used the
recommended procedure and fit the sources in the timeline \citep{Pearson2014}. This method does not require
aperture corrections. T25 was considered as an upper limit at 500\,$\mu$m in \citet{Ribas2013}, but the
  procedure in this study successfully detected it in this band. The obtained flux value does not
  conflict with the previous upper limit in \citet{Ribas2013}. Given the better method used here and the
  flux consistency with the overall shape of the SED, we chose to include it. Conversely, the timeline fitter
does not detect CS~Cha at 500\,$\mu$m probably due to the strong background, and therefore we do not include this
wavelength in its SED. We applied color correction factors corresponding to black-body emission assuming SPIRE
bands trace the Rayleigh-Jeans regime (0.945, 0.948, 0.943 for SPIRE 250, 350, and 500 bands, respectively, see
SPIRE Handbook Version 2.5). Finally, we adopted conservative photometric uncertainties of 20\,\% to account for
different effects (i.e., absolute flux calibration, background estimation). The obtained \emph{Herschel} photometry
is provided in Table~\ref{tab:fluxes}.

\begin{table*}
  \caption{\emph{Herschel} fluxes of the modeled (pre-)transitional disks in this study. Ellipsis indicate non-detected
  sources.}\label{tab:fluxes}
  \begin{center}
    {\footnotesize
    \begin{tabular}{l c c c c c c}
      \hline
      \hline\rule{0mm}{3mm}Name & F70 & F100 & F160 & F250 & F350 & F500 \\
      & (Jy) & (Jy) & (Jy) & (Jy) & (Jy) & (Jy)\\
      \hline
      SZ Cha & 4.01$\pm$0.80	 & 3.74$\pm$0.75	 & 3.56$\pm$0.71	 & 2.53$\pm$0.51	 & 1.85$\pm$0.37	 & 1.02$\pm$0.20	 \\
      CS Cha & 3.20$\pm$0.64	 & 2.88$\pm$0.58	 & 2.27$\pm$0.45	 & 1.31$\pm$0.26	 & 1.04$\pm$0.21	 & \ldots \\
      T25 & 0.53$\pm$0.11	 & \ldots & 0.38$\pm$0.08	 & 0.25$\pm$0.05	 & 0.17$\pm$0.03	 &
                                                                                                                   0.06$\pm$0.01	\\
      \hline
    \end{tabular}
  }
  \end{center}
\end{table*}

\subsection{Near/mid IR photometric data of transitional disks}\label{sec:nearIR}

In \citet{Ribas2013}, we compiled photometry from several surveys and catalogs to build well-sampled SEDs of the
TDs in the region. However, the aim of this paper is to model these SEDs, and hence we only select non-redundant
photometric data. We therefore chose to include the following bands in the near/mid-IR: 2MASS \textit{J},
\textit{H}, and \textit{Ks}, and IRAC1 and IRAC2 bands. This selection provides a nice coverage of the 1-6\,$\mu$m
regime, key to separate (pre-)TDs from TDs \citep{Espaillat2010}. It also avoids redundancy (including several data
in a small wavelength domain), which could give excessive weight to certain parts of the SED in the model fitting
process. Typical photometric uncertainties of these measurements are below 5\,\%, but given the main scope of this
paper, possible IR variability of the sources should also be considered to derive proper uncertainties in the
physical parameters \citep{Muzerolle2009_variability}. To account for these two effects (photometric uncertainties
and variability), we chose to set uncertainties to be a 10\,\% of the observed fluxes.

Finally, all photometric points were derredened using the corresponding $A_V$  (see
Table~\ref{tab:stellar_parameters}) and the extinction law in \citet{Indebetouw2005}.

\subsection{IRS spectra of transitional disks}\label{sec:IRS}

\begin{figure}
\centering
  \includegraphics[width=\hsize]{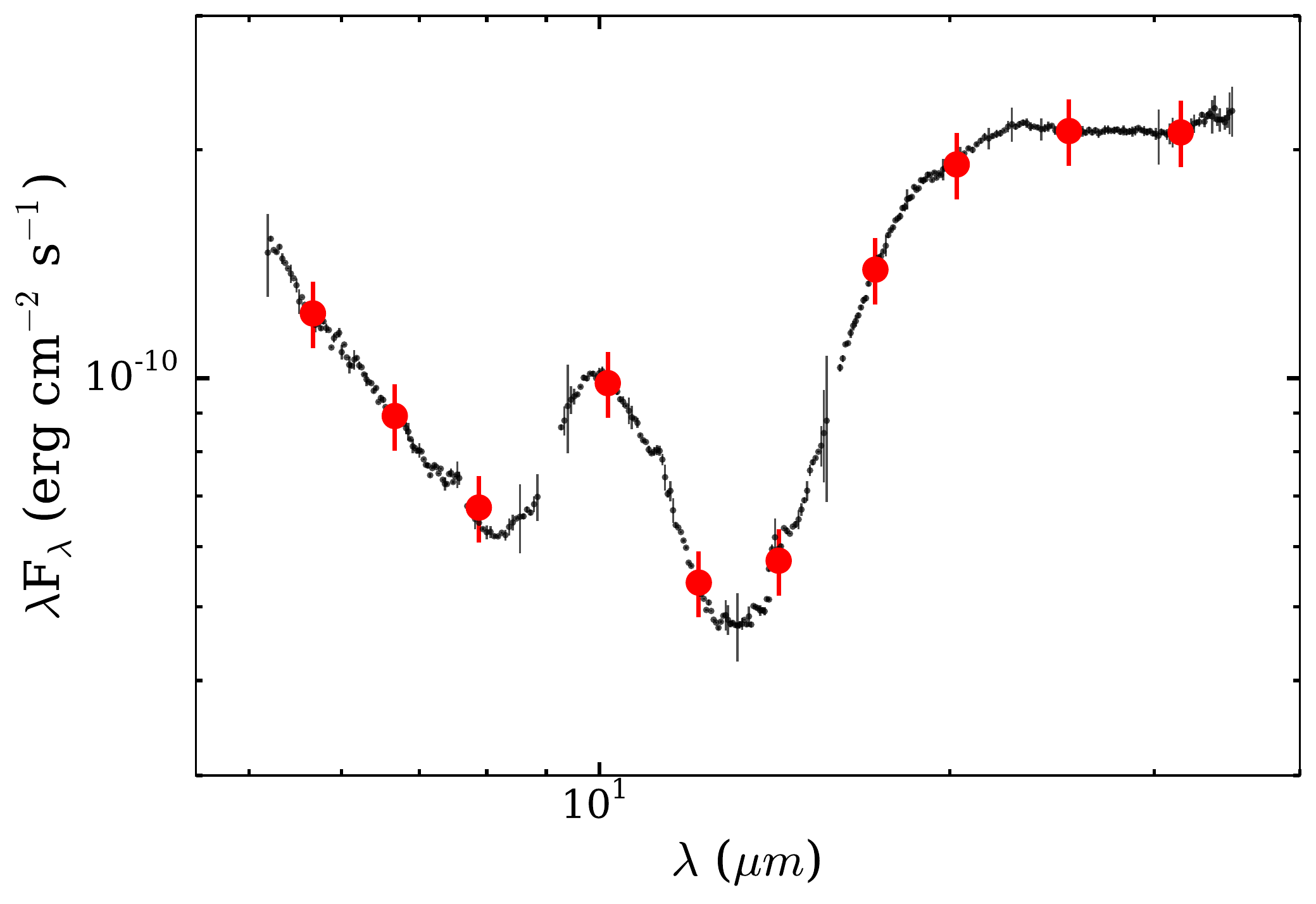}
  \caption[Binned IRS spectrum of SZ~Cha]{Derredened IRS spectrum of SZ Cha. Black dots show the CASSIS spectrum of
    this source (after bad pixels rejection) with the corresponding error bars. The binned spectrum and assumed
    uncertainties are shown as red, larger dots. It properly traces the shape of original data including the
    silicate feature at 10\,$\mu$m.}\label{fig:IRS_spectra}
 \end{figure}

We retrieved low resolution IRS spectra from the Cornell Atlas of Spitzer/IRS Sources (CASSIS) database
\citep{CASSIS} for the (pre-)TDs disks in our study. CASSIS provides optimally extracted IRS spectra, and is
well suited for our purposes. For these spectra, we first separated the optimal zones of the IRS spectra (7 to
14\,$\mu$m and 20.5 to 35\,$\mu$m for the first order, $<$\,20.5\,$\mu$m for second one), and rejected bad pixels
(e.g. negative or NaN values). As a compromise between estimating monochromatic fluxes required for model fitting
(see Sect.\ref{sec:modeling}) while reducing the impact of possible artifacts in the spectra, we chose to bin them
in ten equally spaced wavelengths throughout the spectra coverage, and estimate the fluxes for each of them as the
mean value of ten pixels centered around each corresponding wavelength. We checked this procedure to produce nice
sampling of the IRS spectra (see Fig.~\ref{fig:IRS_spectra} for an example), while being a good compromise for the
SED fitting: a whole IRS spectra typically contains 300-400 good pixels, and fitting them all would put most of the
weight on the IRS spectra itself. By reducing its contribution to a comparable number to that of photometric data
($\sim\,10$) we ensure that all parts in the SED contribute to the fitting procedure in a similar
manner. Additionally, this binning choice allows to encapsulate the basics of the silicate feature (i.e. its
presence and strength). As in Sect.~\ref{sec:nearIR}, we assigned 10\,\% uncertainties to the binned data, a
typical variability value for these disks \citep{Espaillat2011}.

\section{Modeling}\label{sec:modeling}

We aimed at modeling the selected targets and quantifying the impact of adding photometric \emph{Herschel} data
to this process. For this reason, we used two different datasets for each object. The first dataset comprises the
available data from 2MASS, IRAC1/IRAC2, and the binned IRS spectra. The second dataset also includes the
\emph{Herschel} photometry.

We used the MCFOST software \citep{MCFOST, MCFOST2} version 2.19 to model these disks. MCFOST is a Monte Carlo-based
raytracing code which generates synthetic SEDs and images of circumstellar disks. First, it produces temperature
and density maps of the disk using the provided stellar and disk parameters. In this case, we used 10$^7$ photons
in this step (enough to produce smooth and well-sampled temperature maps of the disks). After this step, a list of
wavelengths is provided for MCFOST to calculate the corresponding synthetic monochromatic fluxes. We required
2\,000 photons to be received for each wavelength, corresponding to noise levels of 2-3\,\% in the flux estimates
and well below the assumed observational uncertainties (10-20\,\%).

Our models include seven free parameters: disk dust mass ($M_{dust}$), inner and outer radii
($R_{in}$, $R_{out}$), scale height at 100\,AU (H$_{100}$), flaring index ($h$), surface density
exponent ($p$), and the maximum grain size ($a_{max}$). Given the complex structures of
circumstellar disks, there are several degeneracies and dependencies between these parameters, and
some may even be totally unconstrained with the available data. We did not attempt to fit the
mineralogy of the disks. Instead, we assumed typical astronomical silicate compositions, and fixed
the minimum grain size to 0.01\,$\mu$m.  A more in-depth study of the mineralogy of these disks
would add an important source of complexity to modeling, and we preferred not to include it in our
comparative analysis. We chose a power-law index for the surface density profile. More complex
structures such as tapered-edge profiles could also be used \citep[e.g.][]{Lynden-Bell1974}, but
direct high-resolution observations are required to actually trace the mass distribution in the
disk. Following \citet{Espaillat2011}, we also fixed the inclination of all disks to
60\,\degr. Although this inclination is somewhat arbitrary, none of these objects show signatures of
high-inclination (e.g., silicate features in absorption or under-luminous photospheres). Moreover,
for wavelengths $>$\,13\,$\mu$m, the mid-IR continuum is almost insensitive to this effect unless
the disk is very close to edge-on \citep{Furlan2006,Dalessio2006}.

Among the objects in the sample, SZ~Cha has NIR excess over the photospheric emission. This feature
is characteristic of pre-TDs, sources with an optically thick inner disk, separated from the outer
disk by a gap in the radial dust distribution \citep[e.g.,][]{Espaillat2007a}. Additionally, CS Cha
has no NIR excess but a prominent silicate emission feature at 10\,$\mu$m, indicating the presence
of optically thin dust in its inner hole. The inner disks of these objects have already been modeled
in detail \citep[e.g.,][]{Espaillat2007b,Kim2009,Manoj2011} and we do not attempt to fit them:
instead, we adopted the parameter results from these previous studies to reproduce the NIR SED
shape. The inner disks remained fixed during the fitting process. This may have an impact in our
final results, as discussed later in the paper (see Sect.~\ref{sec:caveats}).

\subsection{Methodology}

We adopted a Bayesian approach to properly derive confidence intervals for the outer disk parameters. The usage of
Bayesian techniques has increased significantly in Astrophysics during the past years, and we do not intend to
explain them in detail. Instead, we refer the interested reader to introductory works such as
\citet{Trotta2008}. Also, this technique has already been applied for modeling circumstellar disks with
\emph{Herschel} data \citep[e.g.][]{Cieza2011TCha, Harvey2012b, Spezzi2013}, mainly via model grids. Here we
describe the adopted fitting procedure.

Bayesian analysis requires that we assign priors to model parameters. While the selection of restrictive priors may
have a significant effect on the fitting results, priors are also an important tool to force parameters to remain
within certain ranges, avoiding non-physical solutions. We used flat (non-informative) priors for all the
parameters, and constrain them to reasonable values for TDs. The prior ranges used were as follows:

\begin{itemize}

\item $\log{(M_{dust}/M_\odot)}$: from -6 to -2,

\item $R_{in}$: from 1\,AU to $r_{in-out}$,

\item $R_{out}$: from $r_{in-out}$ to 500\,AU,

\item $H_{100}$: from 0.5 to 25\,AU,

\item $h$: from 0.8 to 1.3,

\item $p$: from -2.5 to 1,

\item $\log{(a_{max}/\mu{\rm m})}$: from -1 to 4,

\end{itemize}

where $r_{in-out}$ depends on the target, and is a physically meaningless parameter merely used to
avoid the outer disk becoming smaller than the inner one during the evolution of the MCMC. Based on
previous results \citep{Kim2009,Espaillat2011}, we set $r_{in-out}$ to 30, 40, and 50\,AU for T25,
SZ~Cha, and CS~Cha respectively. $M_{dust}$ and $a_{max}$ can take values within several orders of
magnitude, and hence we chose to explore them in logarithmic scale.

We used a modified version of Markov Chain Monte Carlo methods (MCMC) called {\it ensemble samplers with affine
  invariance} \citep{Goodman2010}. This method uses several ``walkers'' or individual chains to explore the
posterior distributions of parameters, and is especially useful when these distributions have complex forms. We
used a slightly modified version of the implementation by \citet{emcee}, and set the stretch parameter of the walk to
1.5, getting acceptance ratios between 10-50\% (a good compromise between a random walk and discarding most of the
proposed positions in the chain evolution).  In every iteration, the chain comprises 100 walkers. We assume
Gaussian uncertainties for our observations, and used the corresponding likelihood function.

To avoid dependencies with the distance to the objects, we normalize every model to the $J$ band
prior to estimating the likelihood. This should have no impact in our results, since the $J$ band traces
photospheric emission in TDs and does not depend on disk parameters (i.e., all models obtained for a
given object have always the same $J$ flux).

When available, we set the initial position of the chains around previous results in the literature \citep{Kim2009,
  Espaillat2011}. The posterior from MCMCs are only valid once the system has lost memory of their
initial values. This can be quantified using the autocorrelation time of the chains, which gives an estimate of the
required number of iterations to draw independent samples. For every case, we computed the
autocorrelation time for each walker in each parameter, and took the maximum value for conservative purposes. We
then left the system evolve for five autocorrelation times (typically $\sim$\,500 iterations). At this point, the
results are independent of the initial position, and the chain is now sampling the posterior distribution. We then
estimated the posterior function with other five autocorrelation times (i.e., 50\,000 models, the result of the 500
iterations per 100 walkers used). 

\subsection{Model caveats and limitations}\label{sec:caveats}

Simple parametric modeling like the one used in this paper offers several advantages (e.g., we can compute
synthetic SEDs of complex disks without analytic solution), but it also suffers from some caveats that should be kept
in mind. Parametric modeling does not guarantee that a combination of parameters is physically consistent, which
we have tried to attenuate using physically meaningful priors. We have not included more complex features/models,
such as puffed up inner rims \citep[e.g.][]{Dullemond2004,Isella2005}, non-axisymmetric inhomogeneities
\citep[e.g.,][]{Andrews2011, vanderMarel2013}, several radial gaps \citep[as ALMA observations have revealed for
HL~Tau,][]{HLTau}, or fitting for the inner disks \citep[e.g.,][]{Espaillat2010,Espaillat2011} or 
mineralogies. Nonetheless, some of these caveats are likely to have little or no impact in our final results, if
applicable at all. The homogeneous treatment of data and fitting procedure used provides a good understanding of
the value of each dataset, and an adequate frame for comparing the obtained distributions.

\section{Results}\label{sec:results}

\subsection{Fitting results without \emph{Herschel} data} \label{sec:no_herschel}

We first use the dataset without \emph{Herschel} data to explore which parameters can be constrained with NIR/MIR
information. The posterior distributions for the three targets are shown in Fig.~\ref{fig:posteriors}, and the
obtained values in Table~\ref{tab:results}. MCMCs also allow to study degeneracies between different parameters by
plotting the chains in different 2-D projections. The degeneracies are very similar in all cases, as revealed by
the cornerplots in the Appendix (Figs.~\ref{fig:SZ_Cha_CP} to \ref{fig:T25_CP}).

The following conclusions can be drawn from the obtained posterior distributions for the model parameters by
analyzing 5-95\,\% confidence intervals (Fig.~\ref{tab:results}):

\begin{itemize}

\item The inner radius ($R_{in}$) can be constrained within 5-20\,AU for all three targets, corresponding to
  relative uncertainties between 80-120\,\%. This is likely due to the fact that most of the NIR/MIR emission
  arises in the exposed wall, hence probing the location of this parameter.

\item NIR/MIR photometry allows to calculate the scale height ($H_{100}$) with uncertainties within 10\,AUs. This is
  expected, since different scale heights modify the amount of stellar flux intercepted by the disk, changing the emission from
the inner disk 

\item The dust mass ($M_{dust}$) and the rest of the geometrical parameters ($R_{out}$, $h$, and $p$)
  show little or no constraint at all with NIR/MIR data alone.

\item Special attention should be paid to the two-peak distributions obtained for $a_{max}$. Any observation at a
  given wavelength $\lambda$ is only sensible to emission from grains of size $a \sim \lambda$
  \citep{Draine2006}. If we allow $a_{max}$ to take small enough values ($<10\,\mu$m), it could produce substantial
  changes in the SED and therefore play a role in the modeling, which may explain the double peaked posterior
  distributions. Although much larger grains are generally expected in disks, this dataset is not enough to resolve
  this effect if we allow $a_{max}$ to take small enough values.

\item As expected, several degeneracies appear in all cases, the most obvious being inner radius with scale height,
and scale height with flaring index. In some cases, the inner radius also shows a dependence with the maximum grain
size, in relation with the previous point.

\end{itemize}

\subsection{Fitting results with \emph{Herschel} data} \label{sec:herschel}

\begin{figure*}
  \centering
  \caption{Derredened SEDs for the transitional disks in this study. Photometric data from previous studies are
    shown as blue solid circles, \emph{Herschel} measurements as orange squares. Uncertainties are plotted,
    although in several cases are smaller than symbol sizes. We also show 100 randomly selected models from the
    obtained posterior distributions for each case: blue lines correspond to fitting without \emph{Herschel} data,
    red lines are the resulting models when including \emph{Herschel} photometry. This gives an idea on the total
    uncertainties in the SEDs of the modeled disks.}\label{fig:SEDs}
  \includegraphics[width=0.33\hsize]{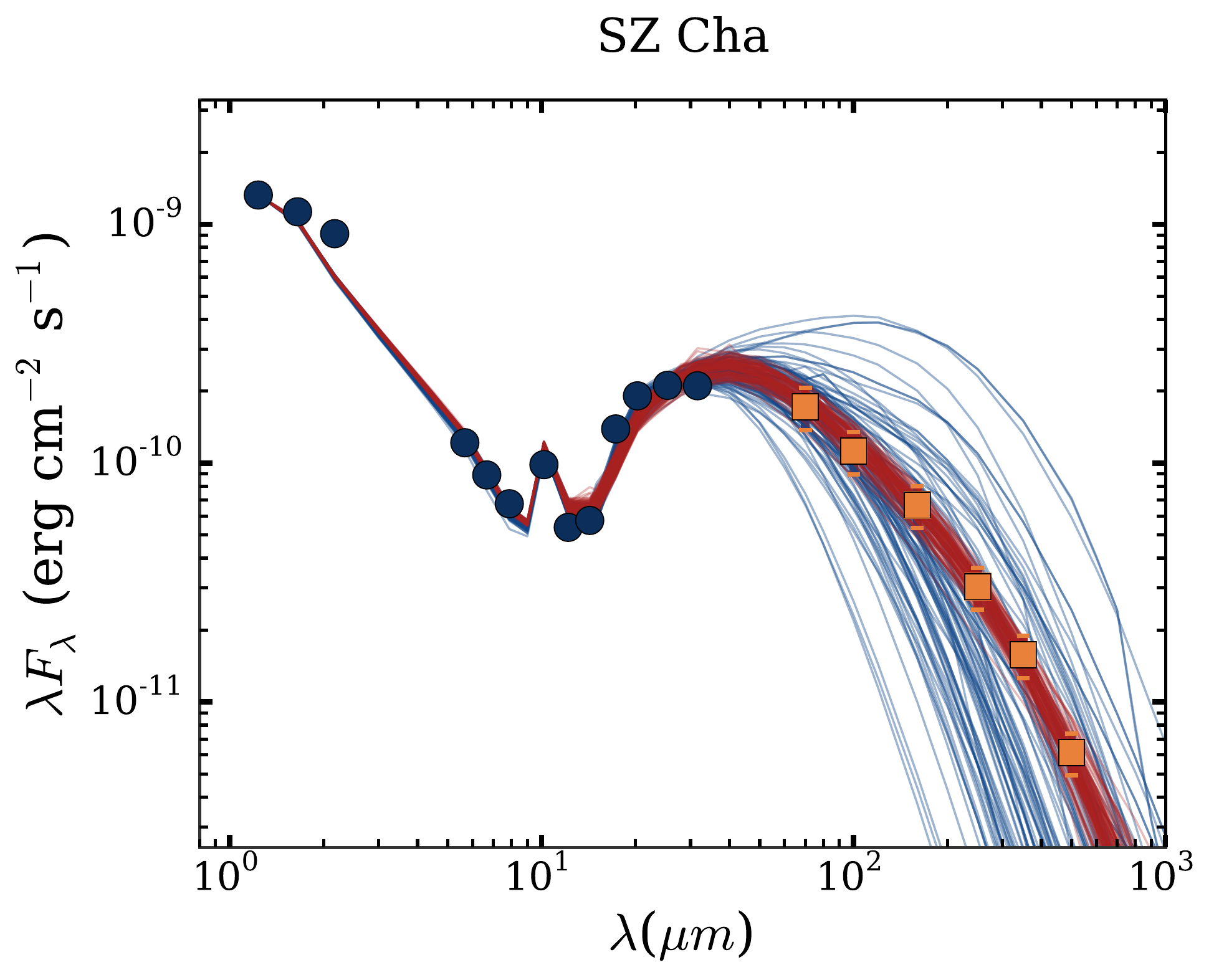}
  \includegraphics[width=0.33\hsize]{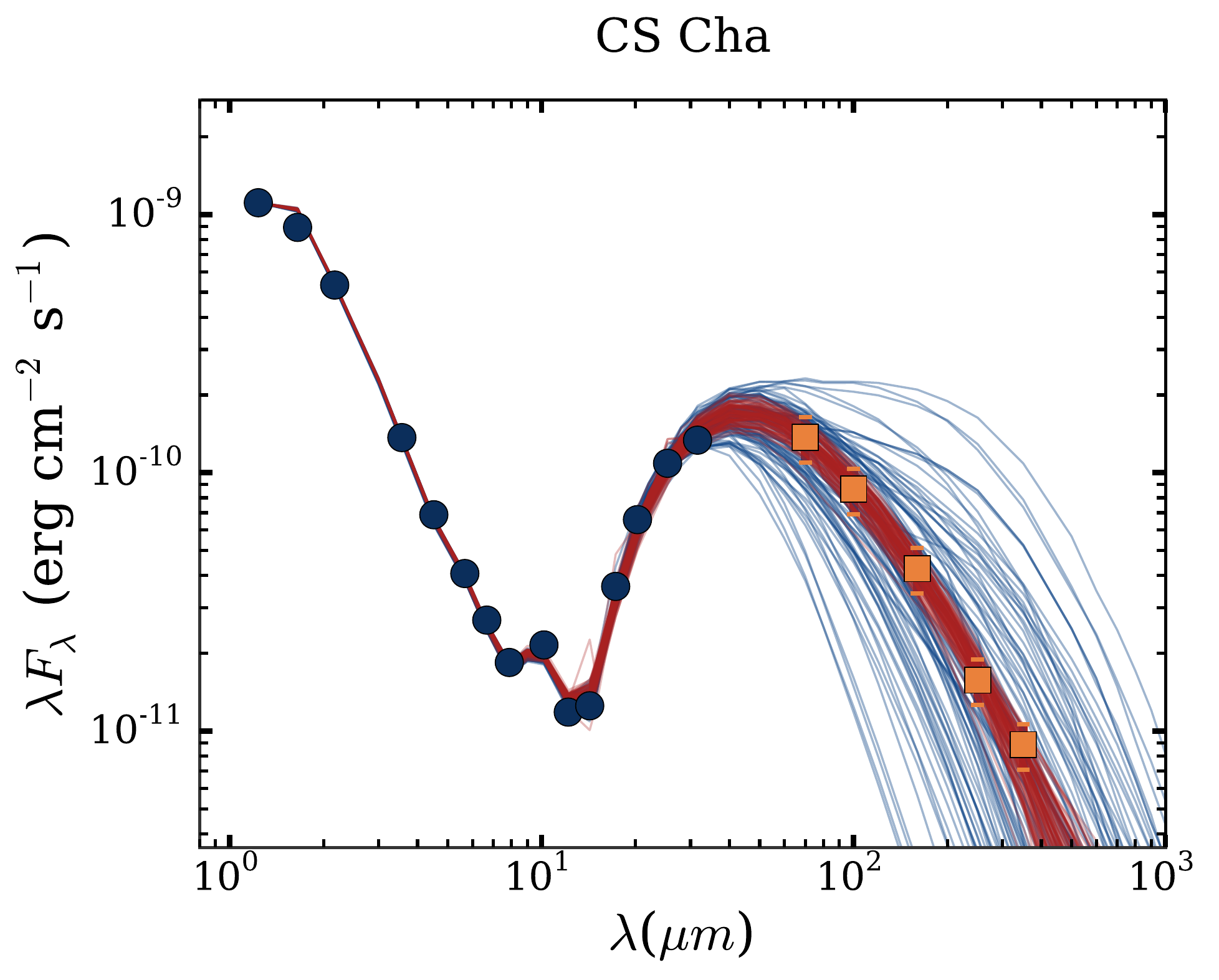}
  \includegraphics[width=0.33\hsize]{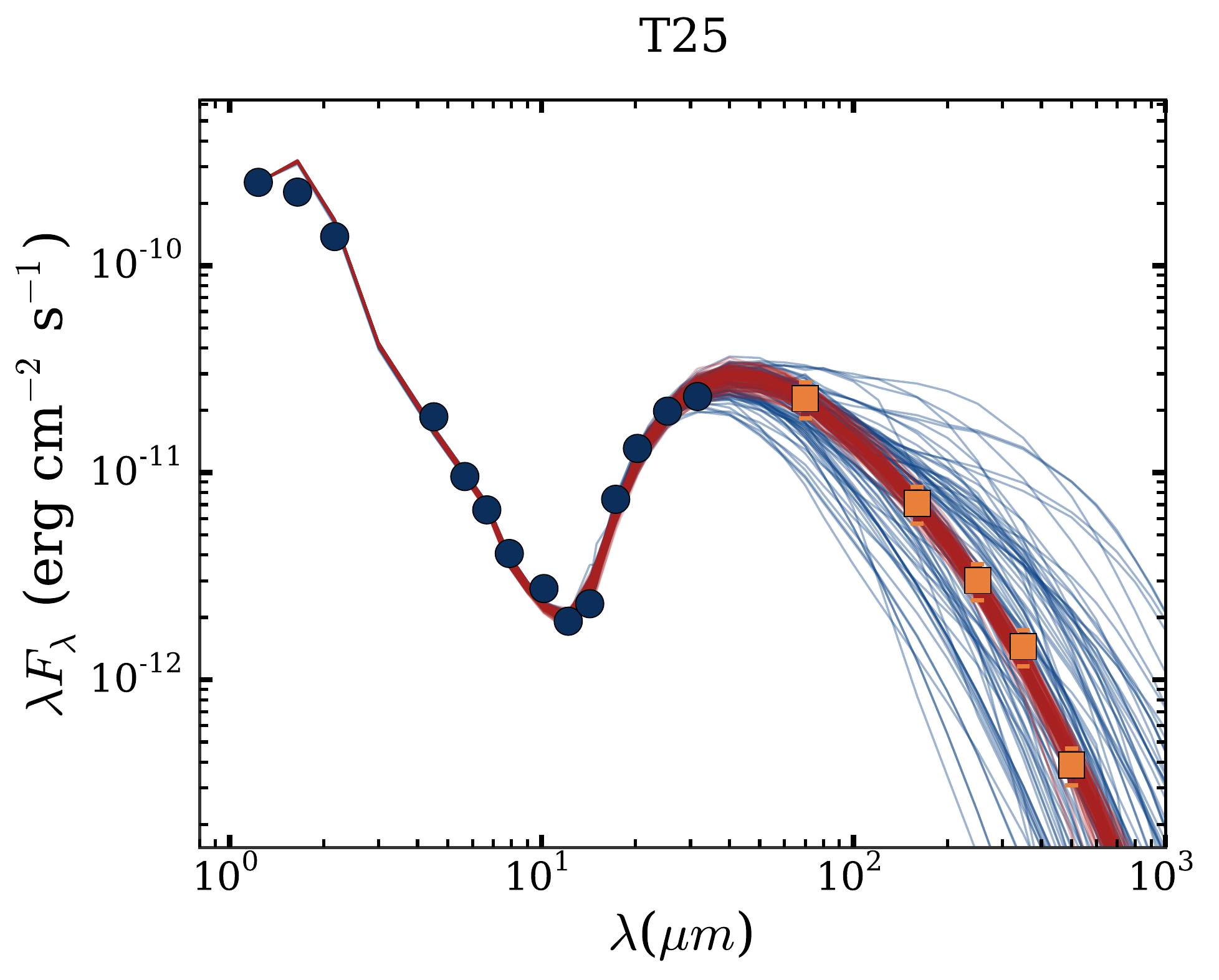}
\end{figure*}

We repeated the modeling procedure including \emph{Herschel} data for the three selected TDs. As in the previous
case, the resulting posterior distributions are shown in Fig.~\ref{fig:posteriors}, and full corner plots in the
Appendix (Figs.~\ref{fig:SZ_Cha_CP} to \ref{fig:T25_CP}). Fig.~\ref{fig:SEDs} show the observed SEDs and modeling
results, and Table~\ref{tab:results} provides the obtained numerical values.

\begin{itemize}

\item Compared to the \emph{Spitzer}-only fit, the addition of \emph{Herschel} data makes an important difference
  for $M_{dust}$ and $R_{in}$. For the latter, the posterior distributions are narrowed down by a factor of two
  with respect to the previous case, with the 5-95\,\% confidence intervals covering 5-10\,AUs, or relative errors
  of 45-60\,\% . For the dust mass, the improvement is substantial, constraining its value within one order of
  magnitude for SZ~Cha and T25, and a broader distribution ($\sim$2\,dex) for CS~Cha, given that it lacks a
  detection at 500\,$\mu$m.

\item The scale height ($H_{100}$) is better constrained with \emph{Herschel} for SZ~Cha and CS~Cha, reducing the
  uncertainties by a factor of $\sim$\,2. For T25, there is no additional improvement, We note that, despite this,
  the combination of $R_{in}$, $h$, and $H_{100}$ yield very similar values of the scale height at the inner
  radius, with improvements in uncertainty below of 1\,AU or less.

\item For the rest of disk geometry parameters, there is no real improvement compared to the \emph{Spitzer}-only
  fit. We do however see marginal evidence of anomalous outer disks in these objects, when combining the preferred
  values of $h$ and $p$, specially for SZ~Cha and CS~Cha. We will discuss this in the following section.

\item \emph{Herschel} data break the two-peak degeneracy in the maximum grain size ($a_{max}$). Although
  they are not enough to provide a real estimate of this value, they inform that this value is very likely larger
  than 100\,$\mu$m in all cases.

\end{itemize}

\begin{figure*}
  \centering
  \caption{Posterior distributions of the free parameters for the
    considered transitional disks. Results without \emph{Herschel} data are shown in blue, those including
    \emph{Herschel} in red.}\label{fig:posteriors}
  \includegraphics[width=\hsize]{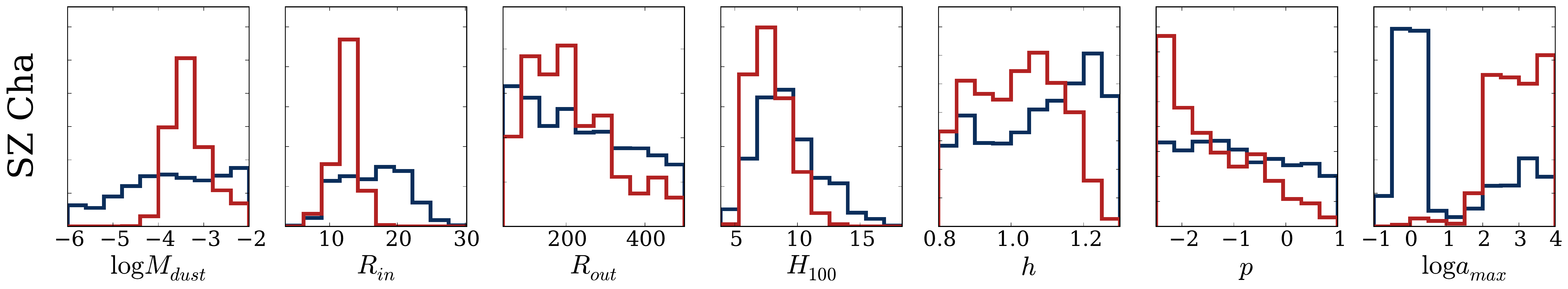} \\
  \includegraphics[width=\hsize]{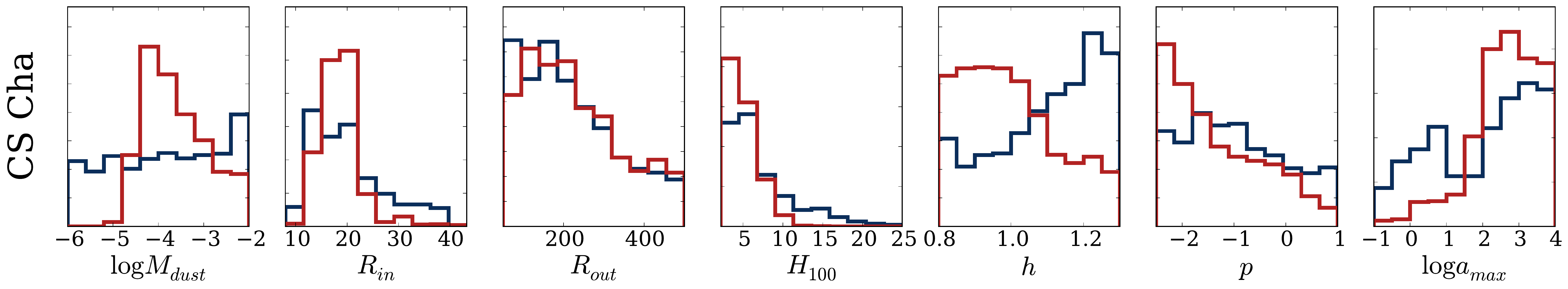} \\
  \includegraphics[width=\hsize]{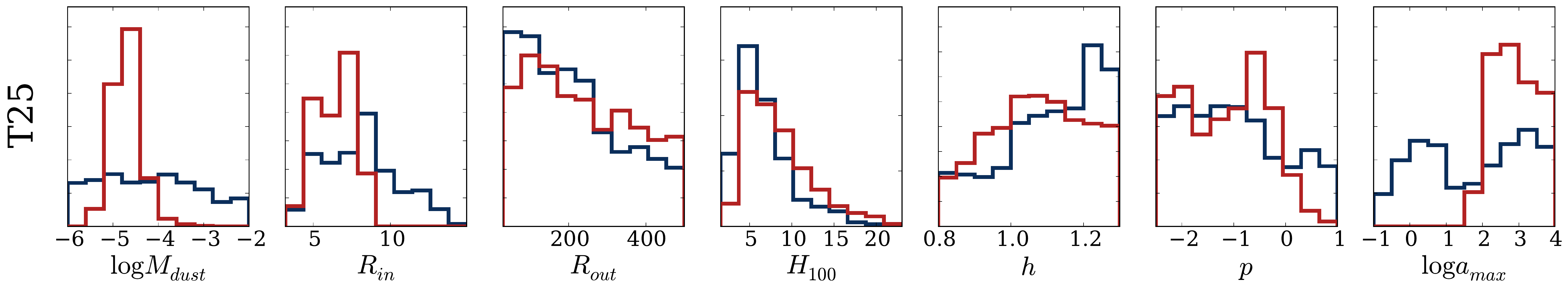}
\end{figure*}

\begin{table*}
  \caption{Results of the MCMC fitting for the seven free
    parameters considered. We tabulate the obtained median value and the 5\,\% and 95\,\% confidence intervals. For
    each column, results obtained without \emph{Herschel} data (left) and with them (right) are
    provided.}\label{tab:results}
  \centering
    \begin{tabular}{l c c c c}
      \hline \hline Object & $\log{M_{dust}}$                            & $R_{in}$                                  & $R_{out}$                                 & $H_{100}$                                  \\
                           & ($\log{M_{\odot}}$)                         & (AU)     & (AU) &  (AU) \\
& no \emph{Herschel} | with \emph{Herschel}& no \emph{Herschel} | with \emph{Herschel}& no \emph{Herschel} | with \emph{Herschel}& no \emph{Herschel} | with \emph{Herschel}\\\hline
      \rule{0pt}{4ex}\vspace{0.3cm}SZ Cha & $-3.6 _{-2.0} ^{+1.5}$ $|$ $-3.4 _{-0.5} ^{+1.0}$ & $17 _{-7} ^{+7}$ $|$ $13 _{-4} ^{+2}$ & $220 _{-170} ^{+240}$ $|$ $200 _{-120} ^{+240}$ & $8.8 _{-3.1} ^{+5.0}$ $|$ $7.6 _{-1.7} ^{+2.9}$ \\
      \vspace{0.3cm}CS Cha & $-3.8 _{-2.0} ^{+1.6}$ $|$ $-3.8 _{-0.8} ^{+1.4}$ & $19 _{-7} ^{+16}$ $|$ $18 _{-5} ^{+6}$ & $200 _{-130} ^{+250}$ $|$ $210 _{-140} ^{+250}$ & $5.8 _{-2.7} ^{+10.4}$ $|$ $4.6 _{-1.3} ^{+4.2}$ \\
      \vspace{0.2cm}T25 & $-4.2 _{-1.6} ^{+2.0}$ $|$ $-4.7 _{-0.5} ^{+0.6}$ & $8.0 _{-3.4} ^{+4.4}$ $|$ $6.6 _{-2.1} ^{+1.6}$ & $190 _{-150} ^{+260}$ $|$ $220 _{-160} ^{+250}$ & $6.0 _{-3.0} ^{+8.1}$ $|$ $7.6 _{-3.8} ^{+9.0}$ \\\hline
\end{tabular}

\vspace{0.5cm}
    \begin{tabular}{l c c c}
      \hline \hline Object & $h$                                             & $p$                                               & $\log{a_{max}}$                                 \\
                           & -                                               & -
                                                                                                                                 & ($\log{\mu m}$)                                 \\
 & no \emph{Herschel} | with \emph{Herschel} & no \emph{Herschel} | with \emph{Herschel}& no \emph{Herschel} | with \emph{Herschel}\\\hline
\rule{0pt}{4ex}\vspace{0.3cm}SZ Cha                     & $1.1 _{-0.3} ^{+0.2}$ $|$ $1.0 _{-0.2} ^{+0.2}$ & $-1.0 _{-1.3} ^{+1.7}$ $|$ $-1.6 _{-0.8} ^{+1.9}$ & $0.2 _{-0.6} ^{+3.5}$ $|$ $2.9 _{-1.1} ^{+1.0}$ \\
\vspace{0.3cm}CS Cha                     & $1.1 _{-0.3} ^{+0.2}$ $|$ $1.0 _{-0.2} ^{+0.3}$ & $-1.1 _{-1.3} ^{+1.8}$ $|$ $-1.6 _{-0.9} ^{+1.9}$ & $2.4 _{-2.8} ^{+1.5}$ $|$ $2.7 _{-2.1} ^{+1.1}$ \\
\vspace{0.2cm}T25                        & $1.1 _{-0.3} ^{+0.1}$ $|$ $1.1 _{-0.2} ^{+0.2}$ & $-1.0 _{-1.3} ^{+1.8}$ $|$ $-1.1 _{-1.2} ^{+1.1}$ & $1.8 _{-2.3} ^{+2.0}$ $|$ $2.9 _{-0.9} ^{+1.0}$ \\\hline

\end{tabular}
   \end{table*}

\section{Discussion}\label{sec:discussion}

\subsection{Masses and inner radii}

The estimate of two free parameters in our models have been found to improve significantly when including
\emph{Herschel} data: the disk dust mass and its inner radius.

The mass of the disk is one of the most important parameters for planet formation. It determines the
available reservoir to build up planets and for accretion on the central star, and can even modify
the planet formation mechanism \citep[the disk instability scenario requires high $M_{disk}/M_*$
values, e.g.,][]{Lodato2005}. Although the bulk of mass in protoplanetary disks is in gaseous form,
photometric IR data are only sensitive to dust emission, which are efficient radiation absorbers and
emitters. Hence, only the dust mass can be (partially) constrained with the presented data. A total
disk mass estimate requires either gas mass measurements \citep[e.g., $^{13}$CO,
C$^{18}$O,][]{Panic2008} or assumptions on the gas-to-dust ratio. Since the former is only available
for a few disks, most authors assume a typical gas-to-dust ratio of 100. We adopt this value for
comparison with previous studies, but are aware that such an assumption may not hold true in many
cases, since the gas-to-dust ratio is likely to change with time and from one source to another
\citep{Thi2010, Thi2014} . Regardless of this, our results show that \emph{Herschel} data can be
used to constrain the mass of dust in these (pre-)TDs within $\sim$\,one order of magnitude for the
5-95\,\% confidence interval, which is a tremendous improvement with respect to MIR photometry only
and opens the exciting possibility of studying this parameter for the large number of sources
observed with \emph{Herschel} and missing mm observations.  As expected and illustrated by the case
of CS~Cha, the longest wavelengths (SPIRE 500\,$\mu$m) are the most important ones to constrain the
mass, as the disk becomes progressively optically thin at longer wavelengths.

The inner radii of the disks are also significantly better constrained with \emph{Herschel} data,
decreasing previous uncertainties by a factor of two. This improvement arises from PACS data, which
narrow down the location of the peak emission, directly related with the illuminated inner
wall. Using unbinned IRS spectra could also provide even better estimates for these values.
Nevertheless, our results results show that at least some information about this parameter is
contained in \emph{Herschel} data.

We also compare the obtained values with previous estimates in the literature. There are four main
studies which can be used for this purpose: \citet{Kim2009}, \citet{Espaillat2011},
\citet{Ubach2012}, and \citet{Rodgers-Lee2014}. \citet{Kim2009} presented detailed modeling of the
IRS spectra of TDs in Chamaeleon~I with an analytic model, including emission from the optically
thin disk and wall and emission from the outer disk treated as a blackbody. In
\citet{Espaillat2011}, the authors used a more complex irradiated disk model \citep{Dalessio2006}
including shadowing of the outer disk by the inner disk \citep{Espaillat2010} to analyze variability
in the IRS spectra of several TDs. \citet{Ubach2012} presented 3 and 7\,mm interferometric
measurements of SZ~Cha and CS~Cha, providing disk mass estimates. More recently,
\citet{Rodgers-Lee2014} performed a multiwavelength study of Chamaeleon~I including \emph{Herschel}
data from \citet{Winston2012}, and analyzed TDs in the region with a physical disk model
\citep{Beckwith1990}.,

\begin{itemize}

\item Mass values computed with two different methods are available for all the targets: via mm data
  \citep{Kim2009, Ubach2012, Rodgers-Lee2014} and via the accretion-to-viscosity ratio
  \citep{Espaillat2011} following \citet{Dalessio1998}. Our results are in very good agreement with
  these previous values, and match within a factor of two for most cases except for the mass value
  of CS~Cha in \citet{Espaillat2011}, SZ~Cha in \citet{Ubach2012}, and T25 from \citet{Kim2009}. In
  the former case, a value of 0.3\,$M_{\odot}$ is quoted, more than one order of magnitude larger
  than our estimated median value (0.015\,$M_{\odot}$) but within the 95\,\% confidence interval
  range. Therefore, the two results are consistent within uncertainties. Additionally, the value in
  \citet{Espaillat2011} depends on the disk viscosity, which is usually largely uncertain and could
  account for this difference. In the case of SZ~Cha, our results for the 5-95\,\% interval yields
  values of 1.3$\times10^{-2}$$-$0.4\,$M_\odot$,
  while \citet{Ubach2012} obtained a total disk mass of 9.4$\times10^{-3}$\,$M_\odot$.
  Considering that this measurement is also subject to uncertainties \citep[between a factor of two
  to ten, according to][]{Ubach2012}, then our results match completely within the uncertainty
  range. For T25, \citet{Kim2009} estimate a 0.007\,$M_{\odot}$
  disk mass via 1.3\,mm fluxes from \citet{Henning1993}. Our study yields a disk mass for T25 of
  0.002\,$M_{\odot}$,
  with their value lying just at the border of the corresponding confidence interval. However, as
  noted by \citet{Rodgers-Lee2014}, the 1.3\,mm flux value in \citet{Kim2009} for T25 is an upper
  limits, and therefore their mass estimates should be considered as such, solving the
  discrepancy. \citet{Rodgers-Lee2014} also found that \emph{Herschel} data within the
  160-500\,$\mu$m
  range can be used to estimate disk masses within a factor of 3 without the need of detailed
  modeling. Our results show larger uncertainties ($\sim$
  one-two orders of magnitude for the 5-95\,\% confidence interval), stressing the importance of
  considering other sources of uncertainties (such as disk temperature and composition) to compute
  realistic confidence intervals of model parameters.

\item Disk inner radii estimates are available both in \citet{Kim2009} and
  \citet{Espaillat2011}. Our results using MIR data only are in general good agreement with their
  values, with the discrepancy of CS~Cha. These two works estimated its inner disk radii to be 41
  and 38\,AU, respectively, while we obtain $19_{-7}^{+16}$\,AU
  with similar data (i.e., excluding \emph{Herschel}). Their results fall outside the 5-95\,\%
  confidence intervals derived in this study. Two different effects can explain this apparent
  discrepancy. First, there is no uncertainty estimation in the quoted studies: if we assume their
  results to have similar uncertainties to ours, the resulting distributions would overlap
  significantly and yield consistent values. Additionally, these two works included more complex
  dust compositions, which can modify the grain emissivity and therefore change the location of the
  inner radius. We also note that \citet{Kim2009} estimated a 29\,AU gap for T25, although the
  improved estimate of 18\,AU in \citet{Espaillat2011} is completely consistent with ours.

\end{itemize}

The mass ranges of these TDs are similar to those of Class~II disks in other star-forming regions
\citep[e.g., Ophiuchus, Taurus,][]{Andrews2010b,Andrews2013}, a result already found by
\citet{Andrews2011} for 12 TDs observed with sub-mm interferometry. This is somehow intriguing: if
TDs are an evolved stage of protoplanetary disks, then we would expect them to have significantly
lower masses. In fact, other works found TDs to have masses even higher than those of Class~II
sources \citep{Najita2007, Najita2015}. If that is the case, TDs (at least classical ones, those
with large holes in their dust distribution) could be the evolution of high-mass disks which have
formed multiple giant planets \citep[explaining their gaps,][]{Zhu2011}, and not a general
evolutionary stage for all protoplanetary disks.

We also compare these values with that of the Minimum Mass Solar Nebula
\citep[MMSN,][]{Hayashi1981}, the minimum mass required to form the Solar System. A typical value of
this quantity is $\sim$0.02\,$M_{\odot}$ \citep{Davis2005,Desch2007}. Both SZ~Cha and CS~Cha are
above or close to this value, meaning that despite being in a transitional stage, they still have
enough mass to form a significant number of planets (although this does not guarantee that planet
formation will take place in the future).

\subsection{Anomalous outer disks}

We find a general trend for flaring indexes ($h$) close to $\sim$\,1, slightly smaller than those usually
found in protoplanetary disks \citep[$\sim$\,1.1-1.3, e.g.,][]{Chiang1997, Olofsson2013}. Additionally,
\emph{Herschel} data suggest strongly negative surface density profiles, with no peak at -1, as typically assumed
and found in protoplanetary disks \citep[e.g.,][]{Andrews2009}. Surprisingly, the obtained values are closer to
that of the estimated for the MMSN \citep[i.e. -1.5, -2.2,][]{Hayashi1981,Desch2007}.

These two results are likely accounting for an observed trend in the SEDs of these three targets:
they have a significant amount of excess in the MIR range up to 70-100\,$\mu$m \citep[already hinted
in][]{Cieza2011TCha,Ribas2013}, but their slopes between 250-500\,$\mu$m are bluer than those
  of typical Class~II disks. This was found in \citet{Ribas2013} when comparing the SEDs of
(pre-)TDs in Chamaeleon~I with the median SED of the Chamaeleon~I and II regions. The obtained steep
surface density profiles and flaring indexes reduce the flux at longer wavelengths (SPIRE), and
increase it at shorter wavelengths (20-150\,$\mu$m, IRS, MIPS, and PACS). Low flaring indexes could
arise if significant dust settling towards the mid-plane has already occurred in these disks,
reducing the disk surface exposed to the stellar radiation specially in the outer regions of the
disk. On the other hand, smaller (more negative) surface densities imply that more mass is located
close to the star, leaving a fainter outer disk which will emit poorly in the FIR regime. Combined,
these results suggest that the modeled (pre-)TDs have anomalous outer disks compared to Class~II
objects. The same fact is found for the T~Cha TD using \emph{Herschel} data \citep{Cieza2011TCha}
and in the Lupus region \citep{Bustamante2015}, reinforcing this idea.

We stress that this interpretation is based on weak evidence and a very small sample, and should be
considered with caution: the posterior functions of these parameters are broad and do not discard
canonical values, but simply make them slightly less probable. The hint of this phenomenon arises
from the fact that the three sources under study show this same marginal behavior. The usage of
tapered-edge surface densities profiles \citep[e.g.,][]{Lynden-Bell1974,Hartmann1998} or puffed up
inner rims \citep[e.g.,][]{Dullemond2001} may also help explaining the anomalous SED slopes. Further
evidence for flattened disks can be obtained by combining accretion and [OI]
measurements. \citet{Keane2014} found the \emph{Herschel} [OI] flux of 26 TDs to be $\sim$2 times
fainter than those of full disks, suggesting smaller gas-to-dust ratios compared to Class~II disks,
or smaller flaring indices. If the first scenario is ruled out by detecting significant gas
reservoirs (via accretion signatures), then the flatter disks explanation would be favored. Resolved
ALMA observations of larger samples of (pre-)TDs and full disks will reveal their real gas content
and probe their structure, shedding light on this open issue.

\section{Conclusions}\label{sec:conclusions}

We use \emph{Herschel} photometry of three TDs in the Chamaeleon~I star-forming region to perform
detailed MCMC modeling of their SEDs and study the impact of \emph{Herschel} data in the obtained results. We find
that \emph{Herschel} photometry, specially from the SPIRE instrument, can be used to constrain the dust mass in
disks within one order of magnitude, as shown by the obtained posterior distributions. \emph{Herschel} data can
also help narrowing down the location of the inner radius of the disk. Our results are in good agreement with
previous studies.

For the modeled targets, we find disk masses comparable to those of Class~II sources in other star-forming
regions. Because TDs are likely to represent a more evolved stage of disk evolution, the fact that they do not have
significantly lower masses could suggest that the typical transitional class (i.e., disks with large gaps in their
dust distributions) is the evolutionary outcome of massive Class~II sources, with enough mass to form several giant
planets which may have cleared their inner regions. Additionally, we find marginal hints of some dust settling
and/or stepper surface density profiles in TDs than in protoplanetary disks. However, this result is tentative and
requires further analysis. A larger sample of TDs, combined with gas and accretion measurements as well as resolved
images of these targets could help solving this issue and shed light on the origin of TDs and their real connection
with planets.

Given the importance of disk masses for planet formation theories, the results obtained in this study open exciting
new options to study this parameter for a large number of targets which lack (sub)mm observations but are present
in the \emph{Herschel} Science Archive. Further calibration of these values could also be achieved with more
  precise disk mass measurements from mm observations. Such a large scale study could identify underlying
relations between the stellar properties, disk masses, and the characteristics of planetary systems.

\section*{Acknowledgments}

We thank the referee Simon Casassus for his review and constructive comments, which helped improving
and focusing the paper. This publication has been possible thanks to funding from the National
Science Foundation. The data processing and analysis presented in this paper have made extensive use
of the following open source software programs: \texttt{Python}, \texttt{IPython}, \texttt{NumPy},
\texttt{SciPy}, and \texttt{matplotlib}. This work has also made a significant use of Topcat
\citep[\url{http://www.star.bristol.ac.uk/~mbt/topcat/}][]{Topcat}. We are grateful to the
developers of these softwares for their contributions. This work is based on observations made with
the Spitzer Space Telescope, which is operated by the Jet Propulsion Laboratory, California
Institute of Technology under a contract with NASA. This publication makes use of data products from
the Two Micron All Sky Survey, which is a joint project of the University of Massachusetts and the
Infrared Processing and Analysis Center/California Institute of Technology, funded by the National
Aeronautics and Space Administration and the National Science Foundation.




\bibliographystyle{mnras}
\bibliography{biblio} 



\appendix
 
\section{Cornerplots for the considered transitional disks}
\begin{figure*}
   \centering
   \includegraphics[width=0.9\hsize]{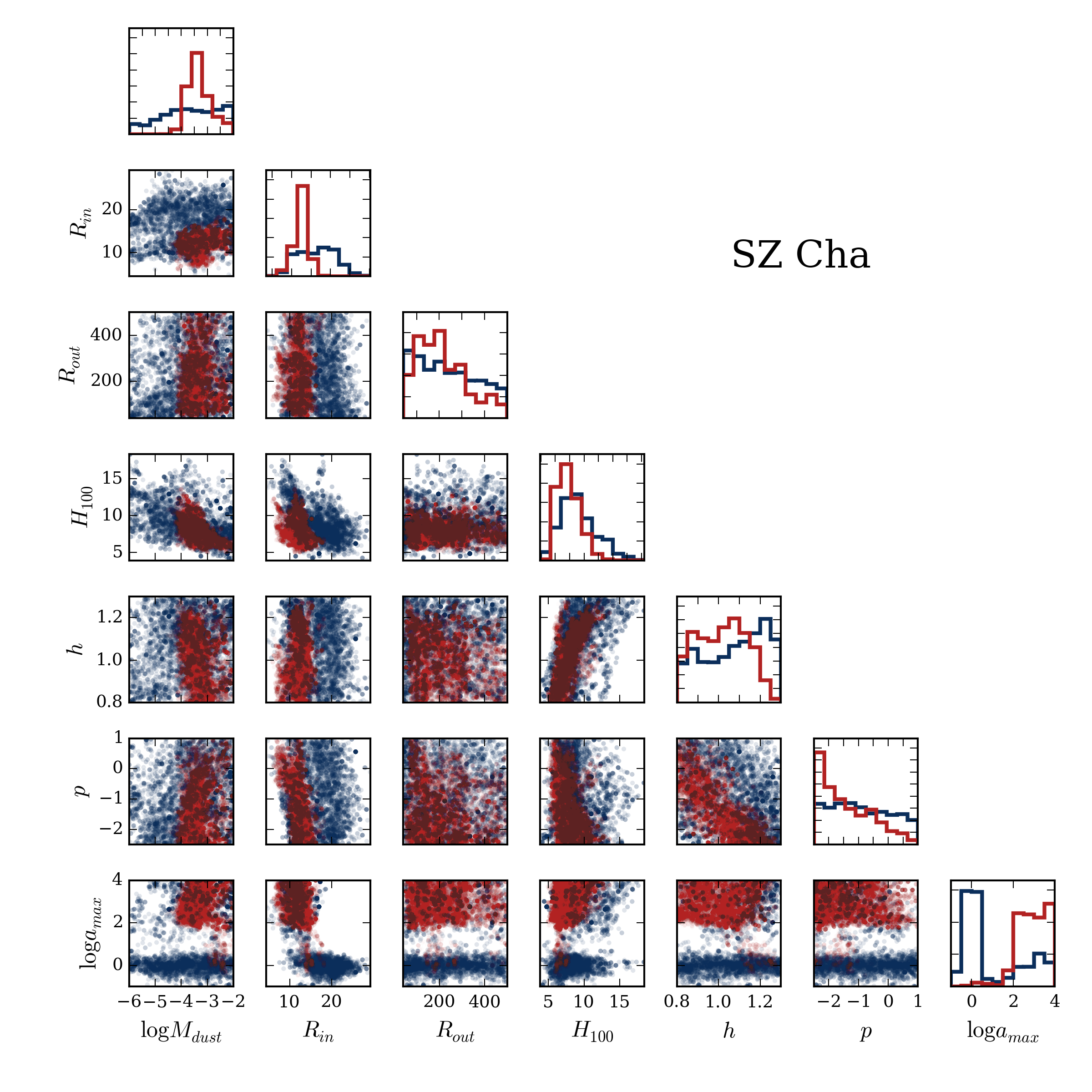}
   \caption{Cornerplot for SZ~Cha. Histograms show the posterior distribution for each free parameter, scatter plots
     display the position of each chain in two parameter spaces to trace degeneracies. The results without
     \emph{Herschel} data are shown in blue, those including \emph{Herschel} in red. }\label{fig:SZ_Cha_CP}
\end{figure*}

\begin{figure*}
  \centering
  \includegraphics[width=\hsize]{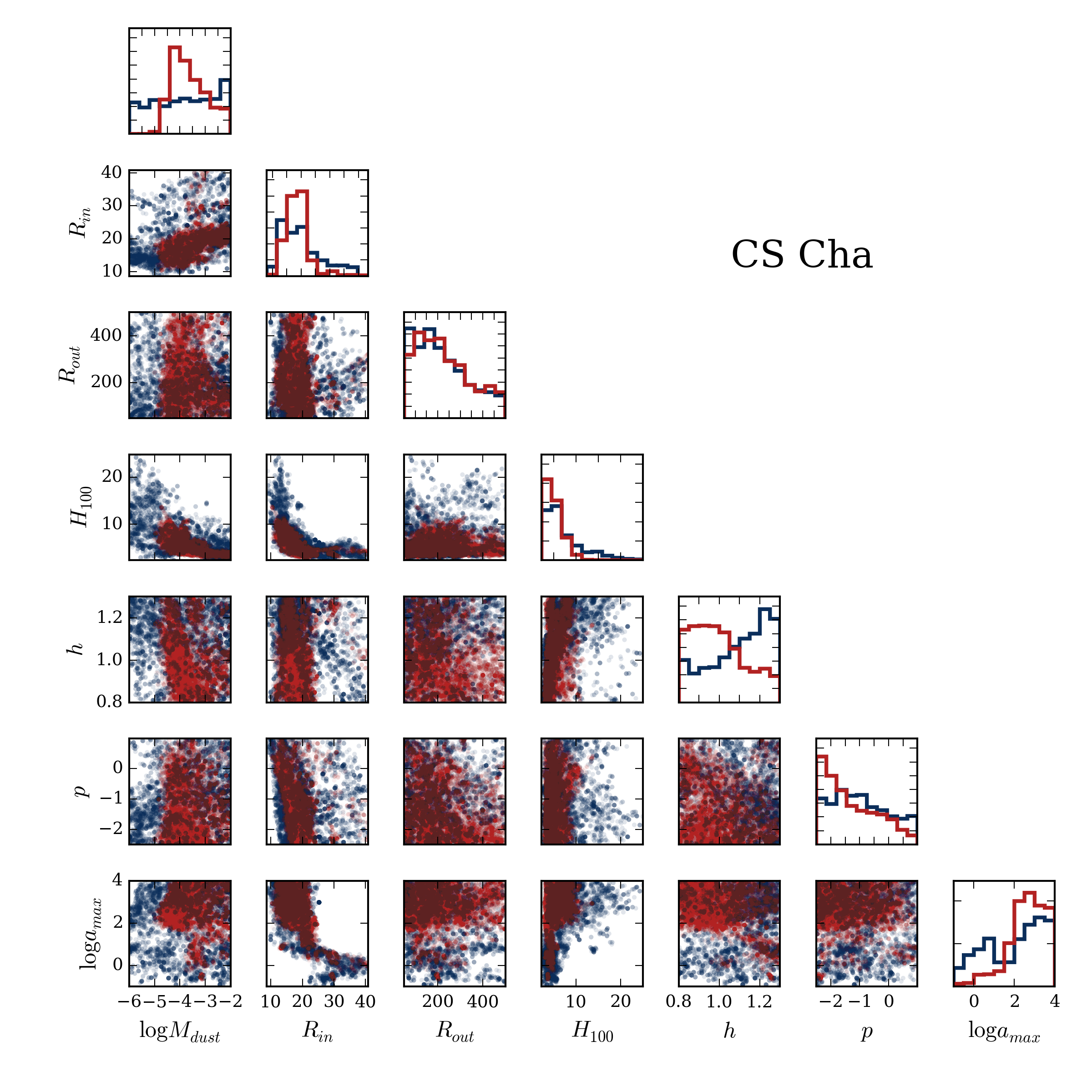}
  \caption{Cornerplot for CS~Cha. Scheme and colors as in Fig.~\ref{fig:SZ_Cha_CP}.}\label{fig:CS_Cha_CP}
\end{figure*}

\begin{figure*}
  \centering
  \includegraphics[width=\hsize]{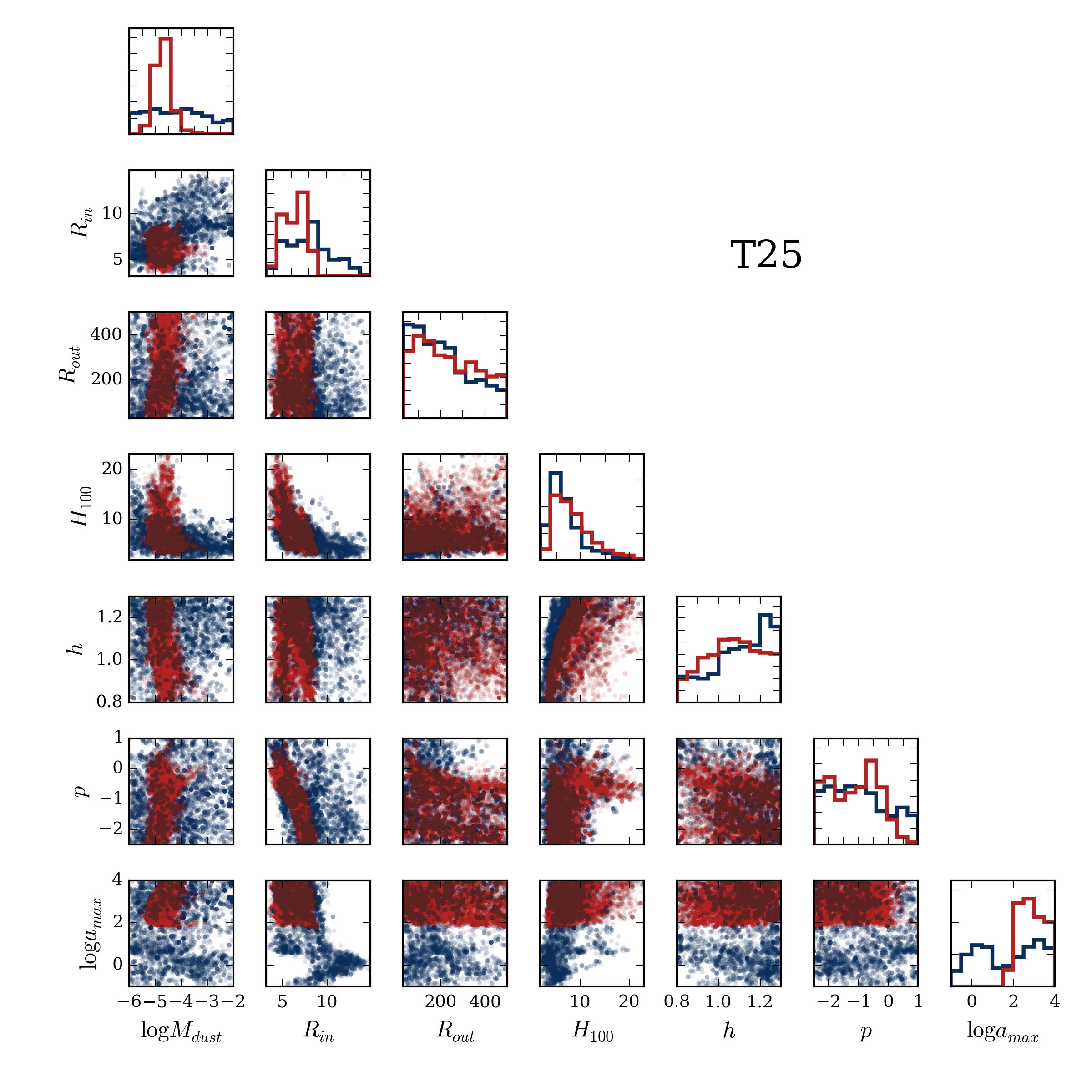}
  \caption{Cornerplot for T25. Scheme and colors as in Fig.~\ref{fig:SZ_Cha_CP}.}\label{fig:T25_CP}
\end{figure*}

In this appendix, we provide the cornerplots obtained with the adopted MCMC procedure
(Figs.~\ref{fig:SZ_Cha_CP} to \ref{fig:T25_CP}).

\label{lastpage}
\end{document}